\documentclass[sigconf,balance=false]{acmart}

\usepackage{popets}

\setcopyright{popets}
\copyrightyear{YYYY}

\acmYear{YYYY}
\acmVolume{YYYY}
\acmNumber{X}
\acmDOI{XXXXXXX.XXXXXXX}
\acmISBN{}
\acmConference{Proceedings on Privacy Enhancing Technologies}
\usepackage[colorinlistoftodos]{todonotes}

\usepackage{booktabs}
\usepackage{tabularx}
\usepackage{pifont}

\usepackage{listings}
\usepackage[T1]{fontenc} 
\usepackage[scaled=0.9]{beramono} 

\usepackage{xcolor}
\usepackage{makecell}

\usepackage{graphicx}
\usepackage{adjustbox}

\usepackage{enumitem}

\usepackage{colortbl}

\usepackage{tikz}
\usetikzlibrary{trees}

\newcommand{\ttpage}{{\small\texttt{page}}}
\newcommand{\serviceWorker}{{\small{\texttt{service\_worker}}}}
\newcommand{\backgroundPage}{{\small{\texttt{background\_page}}}}
\newcommand{\ethAccounts}{{\small\texttt{eth\_accounts}}}
\newcommand{\revokePermissions}{{\small\texttt{wallet\_revokePermissions}}}
\newcommand{\ethRequestAccounts}{{\small\texttt{eth\_requestAccounts}}}
\newcommand{\announceProvider}{{\small\texttt{eip6963:announceProvider}}}
\newcommand{\requestProvider}{{\small\texttt{eip6963:requestProvider}}}

\newcommand{\windowEthereum}{{\small{\texttt{window.ethereum}}}}
\newcommand{\info}{{\small{\texttt{info}}}}
\newcommand{\provider}{{\small{\texttt{provider}}}}

\newcommand{\presentmark}{\(\bullet\)} 
\newcommand{\absentmark}{\(\circ\)} 

\newcommand{\websiteA}{{\small{\texttt{websiteA.com}}}}
\newcommand{\dappX}{{\small{\texttt{dappX.com}}}}

\usepackage{float}
\usepackage{array}
\newcolumntype{x}[1]{>{\centering\arraybackslash\hspace{0pt}}p{#1}}

\newcommand{\TorresOld}{\textit{Torres-2023/100}}
\newcommand{\TorresNew}{\textit{Torres-2025/100}}
\newcommand{\NewEightyFive}{\textit{CWS-10K/85}}

\begin{document}

\title[Privacy Threats of Browser-extension Wallets in the Web3 Ecosystem]{The Masks We (Think We) Wear: Privacy Threats of Browser-Extension Wallets in the Web3 Ecosystem}



\author{Weihong Wang}
\orcid{0000-0003-4296-7271}
\affiliation{%
  \institution{DistriNet, KU Leuven}
  \city{Leuven}
  \state{} 
  \country{Belgium}}
\email{weihong.wang@kuleuven.be}

\author{Yana Dimova}
\orcid{0000-0001-6558-2062}
\affiliation{%
  \institution{DistriNet, KU Leuven}
  \city{Leuven}
  \state{} 
  \country{Belgium}}
\email{yana.dimova@kuleuven.be}

\author{Victor Vansteenkiste}
\orcid{0000-xxxx-xxxx-xxxx}
\affiliation{%
  \institution{DistriNet, KU Leuven}
  \city{Leuven}
  \state{} 
  \country{Belgium}}
\email{victor.vansteenkiste@hotmail.com}

\author{Tom Van Goethem}
\orcid{0000-0001-6846-9081}
\affiliation{%
  \institution{DistriNet, KU Leuven}
  \city{Leuven}
  \state{} 
  \country{Belgium}}
\email{tom.vangoethem@kuleuven.be}

\author{Tom Van Cutsem}
\orcid{0000-0003-4116-4290}
\affiliation{%
  \institution{DistriNet, KU Leuven}
  \city{Leuven}
  \state{} 
  \country{Belgium}}
\email{tom.vancutsem@kuleuven.be}


\renewcommand{\shortauthors}{Wang et al.}

\begin{abstract}

Cryptocurrency wallets are the primary interface for managing pseudonymous blockchain addresses, viewing balances, and interacting with Web3 applications. Although users typically assume that their addresses remain independent of each other unless intentionally revealed, modern wallets routinely communicate with both blockchain infrastructure and decentralized applications (dApps), generating network-side and web-side signals that may undermine this assumption.

In this paper, we identify and formalize five privacy threats that arise directly from wallets interacting with the network and the web browser. Using large-scale dynamic measurements of 85 of the most popular Chrome Web Store browser-extension wallets (representing 35.16 million users), we observe that routine remote procedure call (RPC) operations leak structural links between a user's addresses; that the majority of Ethereum wallets implement permission revocation inconsistently and continue to expose previously revoked addresses across sessions; and that many wallets inject their provider interfaces into cross-origin iframes, enabling passive cross-site tracking beyond dApps and potentially real-world identity deanonymization without user interaction.

Taken together, our results show that these wallet behaviors leak sensitive information that can be used to link multiple addresses to the same user, track wallet users across sessions and sites, and connect their browsing activity to their on-chain wealth.

We discuss practical mitigations and show that many of these threats can be substantially reduced through improved wallet implementation, stronger privacy considerations in ecosystem standards, and stricter controls over provider exposure. Our results highlight the need for standardized, privacy-preserving wallet architectures and provide actionable guidance for strengthening user privacy in the emerging Web3 ecosystem.

\end{abstract}


\maketitle

\section{Introduction}

\begin{figure}[h!]
    \centering
    \includegraphics[width=0.9\columnwidth]{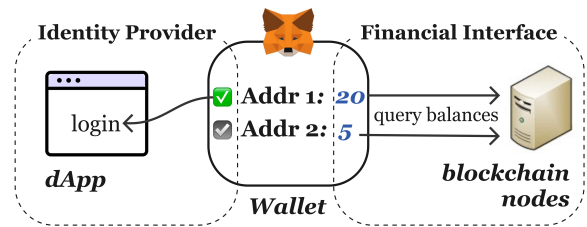}
    \caption{A wallet's two roles: financial interfaces (showing user balances) and identity providers (login).}
    \label{fig:roles}
    \vspace{-1em}
\end{figure}

Browser-extension wallets such as MetaMask form a common entry point to interact with so-called ``Web3'' or ``decentralized applications'' (dApps) in the browser. As illustrated in Figure~\ref{fig:roles}, these wallets typically serve a dual role. First, they operate as a \textbf{financial interface}, querying blockchain nodes (or the wallet's own backend) for balances and other on-chain data associated with the user's addresses. Second, they act as an \textbf{identity provider}: when a dApp requests access, the wallet lets the user select which address to reveal so the dApp can authenticate the session. The wallet thus coordinates the user's interactions with blockchain infrastructure on the one hand, and dApp front-ends on the other hand.

Each blockchain address is a persistent public identifier. Its associated on-chain wealth, complete transaction history, token and NFT holdings, and interactions with smart contracts, is \textbf{visible to anyone who learns the address}. Because a single address reveals such a detailed record of activity, users often spread their actions across multiple addresses to separate identities, reduce risk, or avoid creating a single public on-chain profile. 

This separation is only effective if the addresses remain unlinked. Once two addresses are associated, their histories can be combined, allowing external parties to infer the user's overall asset ownership, behavioral patterns, and financial status, increasing the risk of targeted phishing or other unwanted attention. 

Although many users treat addresses as separate pseudonyms unless they explicitly disclose a connection, prior work~\cite{androulaki2013evaluatingbitcoinprivacy,KapposAnonymityInZcash,victor2020ethereumaddress,sarah2016bitcoinpaymentswithnames} has shown that this assumption does not fully hold, since addresses can often be linked through on-chain graph analysis. 

Our study shows that the privacy risks are even broader than what can be inferred from on-chain activity alone. As the main entry point to Web3, wallets introduce additional privacy risks. Because they mediate both blockchain data queries and identity flows, wallets generate network-level and web-level signals that leak wallet addresses and enable address linkability in previously undocumented ways. While earlier work has shown that wallets leak addresses to external endpoints~\cite{torres2023walletprivacy}, we show that routine wallet behavior, including background network traffic and the exposure of provider interfaces to web pages, can reveal additional sensitive information about users and their addresses. By analyzing both network-side and web-side data flows, we identify five classes of privacy threats:

\begin{itemize}[topsep=5pt]
    \item Network-side \#1: network-level address linkability;
    \item Web-side \#1: wallet fingerprinting;
    \item Web-side \#2: cross-session tracking and address clustering within a single dApp;
    \item Web-side \#3: cross-site tracking and address clustering across multiple dApps;
    \item Web-side \#4: tracking and deanonymization beyond dApps.
\end{itemize}



Our measurements show that these privacy threats are prevalent across the current Web3 ecosystem rather than isolated to a few misconfigured wallets. Across the 85 wallets in our dataset, representing 35.16 million Chrome Web Store users, we observe substantial exposure at both the network and web layers. At the network layer, 17 wallets leak structural linkability signals, affecting 23.0 million users (\textbf{65.4\%}). At the web layer, 36 wallets, covering \textbf{82\%} of the total user base, are detectable through our new fingerprinting vector. More importantly, among these 36 wallets, 22 fail to correctly revoke permissions when users log out of a dApp and continue to expose previously granted but revoked addresses. This persistent exposure can be exploited by third-party trackers to link users across sessions and dApps. Additionally, 23 of these wallets inject their provider interfaces into cross-origin iframes, allowing third-party trackers on non-dApp sites to obtain a user's wallet address without any explicit user interaction. This can link a user's browsing activity to their on-chain wealth and, in some cases, facilitate deanonymization.

Our contributions are as follows:

\begin{itemize}[topsep=3pt]
	\item We perform the first systematic analysis of privacy risks in browser-extension wallets across both the network and web layers, and identify five privacy threats with ecosystem-wide impact. 
	\item We conduct a large-scale empirical study of network traffic and web-side data flows from 85 widely used Chrome Web Store wallet extensions. 
	\item We show that these threats are widespread in practice, and trace two major sources of exposure to incomplete permission revocation and unsafe provider injection into cross-origin iframes.
	\item We propose and discuss mitigations for all five threats, including a script-level access-controlled localStorage design to prevent wallet-based tracking.
\end{itemize}

The remainder of this paper is organized as follows. Section~\ref{sec:bg} provides background on browser-extension wallets. Section~\ref{sec:methodology} details our methodology. Section~\ref{sec:network-threats} defines and evaluates network-side privacy threats, while Section~\ref{sec:web-side} introduces four web-side threats and examines their impact. We discuss mitigation strategies in Section~\ref{mitigation}, related work in Section~\ref{sec:relatedwork} and conclude in Section~\ref{sec:Conclusion}. We release artifacts and a demo in Section~\ref{sec:EthicalPrinciples}, and document our responsible disclosure process in Section~\ref{sec:responsible}.

\section{Background}~\label{sec:bg}

This section introduces how browser-extension wallets function and how decentralized applications interact with them. We describe the primary roles of the wallets, how they are detected by dApps, and how permissions governing these interactions are managed.



\subsection{Roles of Browser-Extension Wallets}~\label{subsec:bg-wallets}

\subsubsection{\textbf{Wallet as a Financial Interface}}

When functioning as a lightweight financial interface, a wallet must routinely fetch on-chain data such as account balances, token metadata, transaction history, or gas prices. For example, to display a balance, a wallet issues an RPC (remote procedure call) request such as: \textit{eth\_getBalance(address)}. Any blockchain node that exposes an RPC endpoint can respond to these queries. 

Because running a full blockchain node is resource-intensive, most wallet developers rely on third-party node providers (e.g. Infura, Alchemy, Ankr, QuickNode) rather than operating their own infrastructure. Some wallets allow users to configure a custom RPC endpoint, but many use a fixed provider chosen by the developers.

\subsubsection{\textbf{Wallet as an Identity Provider}}


A browser-extension wallet can also act as an identity provider for dApps, similar to logging into a website with Google. Google stores the account, and the website receives only the identity information it needs. In Web3, the wallet holds the user’s private keys and manages their blockchain accounts, and the dApp receives the public address that the wallet discloses only after showing a confirmation prompt to the user.

Once the address is provided, the dApp can independently look up public on-chain data (such as balances, history, or token holdings). 
However, any action that moves funds or signs a transaction still requires an explicit confirmation from the user through the wallet interface. 

Wallets store permission and connection state locally within the extension environment, e.g., which dApps have been granted access to which accounts. This local state determines how the wallet responds to future wallet provider API calls. 

\subsection{How dApps Discover Wallets}

For a dApp to request access to a user’s accounts, it must first detect the browser's available wallets. In Ethereum (the most widely used blockchain for dApps), wallets can be detected as follows:

\subsubsection{\textbf{EIP-1193 (Legacy Provider Injection)}} 

Before October 2023, browser-extension wallets generally exposed their Ethereum provider by injecting an EIP\footnote{EIP stands for Ethereum Improvement Proposal, the standardization process for Ethereum interfaces and protocols.}-1193-compliant~\cite{EIP-1193} object into the webpage's global variable \texttt{window.ethereum}. As the ecosystem grew and users started using multiple wallets at the same time, this became problematic. Each wallet injected its own provider object, and the last one to load would overwrite the others. This made it difficult for dApp developers to detect which wallet the user wanted to connect or to let users choose among several installed wallets.


\subsubsection{\textbf{EIP-6963 (Modern Multi-Wallet Discovery)}}

To support the coexistence of multiple wallet extensions, EIP-6963~\cite{EIP-6963} was proposed, which introduced a discovery mechanism in which wallets announce themselves without competing over \windowEthereum{}. Instead of directly overwriting a single global object, wallets dispatch an \announceProvider{} event containing two fields:

\begin{itemize}
    \item \info{}: metadata about the wallet (e.g., name and icon),
    \item \provider{}: the wallet provider object.
\end{itemize}

A dApp listens for these announcements and can present a UI allowing the user to select a wallet. The \requestProvider{} event was introduced in EIP6963, which prompts all installed wallets to re-announce themselves. This enables reliable multi-wallet detection and avoids conflicts caused by legacy provider injection.


\subsection{Wallets Permissions and Revocation}

A wallet must also control which dApps may access the user's wallet addresses. This permission model, standardized in Ethereum through EIP-2255~\cite{EIP-2255}, operates entirely in the browser and governs how a dApp requests access to identity-related information.


\subsubsection{\textbf{EIP-2255 (Wallet Permissions System)}}


EIP-2255 defines a capability-based permission system for wallet-dApp communication. These permission checks occur entirely in the browser between the dApp and the wallet provider.


Each method exposed by the wallet provider to the dApp is classified as either \textit{restricted} or \textit{unrestricted}. A \textit{restricted} method is one that accesses a capability available only after the user has explicitly approved it. An \textit{unrestricted} method does not itself require prior permission, but calling it may trigger a wallet prompt asking the user to grant permission for a restricted method.

\begin{itemize}
    \item \ethAccounts{}: \textbf{restricted}. It returns either an empty array or only the accounts the caller is permitted to access by the user.
    \item \ethRequestAccounts{}: \textbf{unrestricted}. Calling this method triggers a wallet prompt asking the user to approve or reject \textbf{account access for a given dApp}. Once approved, this permission causes later calls to \ethAccounts{} from the dApp to return the user-authorized accounts until the permission is revoked by the user. 

\end{itemize}

\subsubsection{\textbf{How Wallets Store Permission Information}}

Wallets record which addresses a given website origin is allowed to access using a per-origin authorization state internally, typically in the extension-managed storage. When the webpage calls \ethAccounts{}, the wallet checks this record: If permission exists, the wallet returns the previously authorized addresses. If not, it returns an empty array. EIP-2255 does not define expiration, so wallets differ in how long this state persists. 

\subsubsection{\textbf{Lack of a Standard for Permission Revocation}}

While EIP-2255 specifies how permissions are requested and granted, it does \textit{not} standardize any mechanism by which websites can revoke them. As a result, wallet implementations differ substantially: 

\begin{itemize}
    \item MetaMask defines a revocation method in its MIP-2 proposal~\cite{metamask_mip2}. This method, named \revokePermissions{}, allows a dApp to request that the wallet revokes permissions previously granted to that dApp. Some other wallets follow it and also support this method. 
    \item Many wallets expose no revocation API to dApps, so logging out of a dApp does not revoke previously granted permissions. In such cases, permissions may persist until the user manually removes them for each site through wallet-side UI, such as a ``Connected Sites'' or ``Permissions'' screen. 
\end{itemize}

Because no EIP currently defines a uniform revocation workflow, the persistence and removal of permissions remain inconsistent across wallet implementations.





\section{Methodology}\label{sec:methodology}

\subsection{Measurement Goals}

Our study focuses on two adversaries that can observe different aspects of a browser-extension wallet's behavior:  
(1) \textbf{network endpoint servers}, and
(2) \textbf{web-based adversaries} embedded on webpages. Our measurement goals reflect what each adversary model can see. 

\subsubsection{\textbf{Network-Side Measurement Goals}}

Any endpoint the wallet connects to can see all outbound traffic initiated by a wallet extension to that endpoint. Hence, our goal is to measure the information at the network layer, including:

\begin{itemize}
    \item whether these requests included wallet-related information,
    \item which third-party endpoints receive these address-bearing requests,
    \item whether the overall traffic patterns reveal relationships between multiple accounts in the same wallet.
\end{itemize}

\subsubsection{\textbf{Web-Side Measurement Goals}}

A web-based adversary operates through third-party scripts embedded on a webpage. This adversary’s visibility arises from the wallet’s web-facing provider APIs and events. Our goal is to measure what information becomes accessible to such an adversary at different stages of wallet usage:

\begin{itemize}
    \item \textbf{Installed but not connected}: whether the presence of a wallet can be discovered, 
    \item \textbf{Connected}: what provider APIs, events, or address data become available when a wallet connects to a dApp,
    \item \textbf{Revoked}: what residual information remains accessible after the dApp requests permission revocation. 
\end{itemize}


\subsubsection{\textbf{Methodological Structure}}
These goals motivate two complementary measurement frameworks:

\begin{itemize}
    \item a \textbf{Network Request Interceptor} framework, improved from the request-interception approach of Torres et al.~\cite{torres2023walletprivacy},  which detects leakage of addresses in the wallet's outbound network traffic. We also extended it to support our measurement scope.
    \item a \textbf{Web Exposure} framework that observes the wallet provider interface to detect what data becomes observable to third-party scripts throughout different stages.
\end{itemize}

\subsection{Datasets Construction}

Our measurements require three types of datasets: (i) source-code of browser-extension wallets, to characterize network- and web-side behavior across the ecosystem; (ii) decentralized applications (dApps), to observe real-world wallet–dApp interactions; and (iii) third-party analytics and tracking domains.

\subsubsection{\textbf{Torres et al. Wallet Dataset}}

For our reproduction study of Torres et al.'s results, we used their dataset of 100 wallet extensions, which contains the exact versions they analyzed in 2023. We refer to this data set as \TorresOld{}. 

We also downloaded the latest available versions of these same 100 wallets from the Chrome Web Store as of November 2025. We refer to this updated dataset as \TorresNew{}.

\subsubsection{\textbf{\NewEightyFive{} (85 Chrome Web Store (CWS) Wallets with >10K Users)}}

Because more than 25\% of the extensions in the Torres' dataset are no longer available on the Chrome Web Store and the list is outdated, we crawled a fresh list of wallet extensions directly from the store. Using the keywords ``crypto wallet'', ``web3 blockchain wallet'', and ``web3 wallet'', we collected a total of 198 entries and kept only those with more than 10K users on the Chrome Web Store (November 2025). This filtering step produced a dataset of 85 modern wallets, which we refer to as \NewEightyFive{}.

\subsubsection{\textbf{dApp Dataset (30 Popular Ethereum dApps)}}
To understand how real dApps interact with wallets, we collected the 30 most popular Ethereum dApps from DappRadar’s Ethereum category~\cite{dappradar_eth_rankings} as of 14 November 2025. We only included dApps with functional frontends that could connect to an Ethereum-compatible wallet. The goal of this dataset is not exhaustive coverage, but to provide a representative set of real sites through which to observe connection flows, account requests, and revocation behavior.

\subsubsection{\textbf{The List of Analytics Sites}}

We compiled a small list of common third-party analytics and tracking domains using publicly documented providers. The full list of 21 domains used in our classification is provided in Appendix~\ref{appendix:analytics-list}.

\subsection{Network Request Interceptor Framework} \label{subsec:framework-network}

\subsubsection{\textbf{Baseline Framework (Torres et al.)}}


The Puppeteer-based request interceptor framework, developed by Torres et al.,~\footnote{The Puppeteer-based request interceptor framework by Torres et al.: \url{https://github.com/christoftorres/Web3-Privacy/tree/main/framework/request-interceptor}} attaches network listeners to webpage contexts, allowing us to observe the full HTTP and WebSocket request parameters before TLS encryption is applied. However, the framework only monitors requests from browser tabs and does not capture requests originating from Chrome extension components such as background pages (Chrome Manifest V2) or service workers (Chrome Manifest V3)~\cite{chromeMV3,puppeteerDevelop}. Because modern wallet extensions often issue background traffic from these components, such requests appear ``silent'' to the baseline framework even while the extension is actively communicating with external endpoints.


\subsubsection{\textbf{Improved Request Interceptor Framework}}\label{subsec:framework1-improved}

To address the blind spots, we extended the baseline framework to monitor not only browser tabs (Puppeteer targets of type \ttpage{}), but also network activity from extension components (type \serviceWorker{} and \backgroundPage{}). Our modification adds these two extra target types to the interception logic and writes their network events to separate log files. The rest of the data-collection process remains unchanged.

To simulate a real user with multiple addresses in one wallet, the framework records all addresses created during setup, so the addresses can be reused consistently across experiments and analysis.

\subsubsection{\textbf{Account Funding}}

We transferred a small amount of ETH ($\approx$ 1 USD) to the primary two test addresses. This lets us verify that the wallet correctly displays balances, and that it is actively querying external services for blockchain data. Based on this check, we assigned wallets to three categories:

\begin{itemize}
    \item \textbf{\textit{Broken}}: the extension could not be installed, initialized, or opened.
    \item \textbf{\textit{Partially functional}}: the extension loaded and appeared operational, but did not display the correct balance.
    \item \textbf{\textit{Fully functional}}: the extension initialized successfully and displayed the expected balance.
\end{itemize}

Wallets classified as \textit{partially functional} that also produced no outbound traffic to external services were labeled \textbf{\textit{excluded}}, together with the \textit{broken} wallets, omitted from subsequent analysis.

\subsubsection{\textbf{Measurement Procedure}}

Torres et al.'s original study relies on Chrome browser profiles. For each of the 100 wallet extensions, they manually set up the wallet, created a single test account, and saved the resulting browser profile. During testing, Puppeteer reloaded these profiles and ran a routine that randomly clicked up to ten UI elements or stopped after sixty seconds. During this period, the framework recorded all outgoing traffic. 

We followed the same experimental procedure, but performed all measurements using our improved request interceptor framework. During wallet setup, if the extension prompted for telemetry or analytics data collection, we explicitly declined such requests whenever possible. We first ran these experiments on the \TorresOld{} dataset to compare our measurements with the original study and quantify how much traffic the baseline framework missed. We repeated these experiments on the \TorresNew{} and \NewEightyFive{} datasets to evaluate whether these blind spots persist in current wallet versions.

Specifically, we applied the multi-account setup only to the \NewEightyFive{} dataset. Each wallet was initiated with three accounts, and their addresses were recorded. The subsequent traffic-capture procedure was identical to the single-account measurements.





\subsection{Web Exposure Framework}~\label{subsec:framework-permission}

\subsubsection{\textbf{Framework Design}}


We developed a Playwright-based framework to evaluate wallet discovery via EIP-6963 events and the implementation of permission requests, revocation, and reconnection by wallets.
The framework performs four actions for each wallet:

\begin{enumerate}
    \item \textbf{Discover}: discovers a wallet,
    \item \textbf{Connect}: triggers a wallet connection request from dApp,
    \item \textbf{Disconnect}: requests permission revocation from dApp,
    \item \textbf{Evaluate}: checks what address information the dApp can still access afterward.
\end{enumerate}

To create a controlled testing environment, we built a minimal test dApp with only two buttons: a \textbf{Connect} button to trigger \ethRequestAccounts{} after a wallet is discovered, and a \textbf{Disconnect} button to trigger \revokePermissions{}.
During each test, we instrumented the page to record all wallet-dApp permission interactions:

\begin{itemize}
    \item the wallet provider API calls (\ethRequestAccounts{}, \\\ethAccounts{}, \revokePermissions{}, etc.),
    \item the address information accessible by the dApp through calling \ethAccounts{} after disconnect.
\end{itemize}


We also applied the same instrumentation to 30 real Ethereum dApps to understand how modern sites behave in practice. For these real-world dApps, we additionally recorded: 
 
\begin{itemize}
    \item localStorage and cookies modifications,
    \item all outbound dApp requests to external services.
\end{itemize}

These additional real-world samples help reveal broader dApp interaction patterns in wallet discovery and connection.

\subsubsection{\textbf{Measurement Procedure}}

Before running experiments, each wallet in the \NewEightyFive{} dataset was set up once to create a clean browser profile. Our framework supports partial automation for this task, for example, detecting input fields for seed phrases or passwords, but many wallets present non-standard onboarding flows. Because automation is not the focus of this work, we use automated steps when they succeed and fall back to manual interaction when needed.
Each experiment starts from a \textit{copy} of the corresponding clean profile so that the wallet always begins in a fresh state, with no existing permissions or prior connections.
For each wallet-dApp interaction, the framework performs the following steps: 

\paragraph{\textbf{Detecting:}}
We first attempt to detect wallets using the EIP-6963 discovery workflow by dispatching \requestProvider{} and collecting the provider's name.

\paragraph{\textbf{Connecting: }}
\begin{enumerate}[topsep=3pt]
    \item Unlock the wallet.
    \item Load the testing dApp.
    \item Identify and click a ``Connect'' button on the page.
    \item Wait for the wallet’s connection prompt to appear.
    \item (Inside the wallet UI) Select the test account.
    \item (Inside the wallet UI) Appprove the connection request using the wallet’s confirmation interface.
\end{enumerate}

\paragraph{\textbf{Disconnecting: }}
\begin{enumerate}[topsep=3pt]
    \item Identify and click a ``Disconnect'' button on the page.
    \item Trigger the \revokePermissions{} request from the dApp.
\end{enumerate}

Because wallet and dApp interfaces for these steps vary widely, some interactions can be handled automatically by the framework while others require manual clicking.

After the disconnect step, the framework reloads the page and queries \ethAccounts{} to record what information remains accessible to page-executed scripts. The returned value is recorded as the post-revocation state.
The same procedure is applied both to our minimal testing dApp and to the 30 real-world Ethereum dApps.

\subsection{Ethical Principles}
All experiments were conducted using test accounts and isolated browser profiles. No real users or personal data were ever collected, observed, or shared.

\section{Network-Side Threats and Results}\label{sec:network-threats}

This section examines the privacy risks that arise from wallet-initiated network traffic. We begin by defining the network-side threat model and the information available to adversaries through standard RPC (Remote Procedure Call) communication. We then describe the resulting privacy threat and present the empirical results observed across the three datasets using our improved request-interceptor framework.

\subsection{Threat Model} \label{subsub:threat-model:external-network-adversary}

The adversary does not break cryptography, compromise private keys, or interfere with normal wallet functionality, but exploits information gained from standard wallet-initiated network requests.

We consider as the network-side adversary any \textbf{external service endpoint} that receives the wallet’s HTTPS requests during normal operation. These endpoints fall into two categories: (1) \textbf{wallet vendor–operated backends}, which handle or forward the requests to other services; and (2) \textbf{non-vendor-operated domains}, including node providers and RPC endpoints (e.g., Infura, Alchemy, Etherscan), which process blockchain state queries from wallets (see Appendix~\ref{appendix:rpc-providers} for representative domains), as well as analytics or telemetry services that collect usage or diagnostic information about wallet activity (see Appendix~\ref{appendix:analytics-list}).

Because these servers receive the decrypted HTTPS request, they have full visibility into the plaintext request body and can observe:

\begin{itemize}
    \item the full request payload, including wallet addresses, 
    \item the ordering and timing of consecutive requests,
    \item client metadata such as IP address and user agent.
\end{itemize}

The adversary responds correctly to queries but may analyze the content and structure of incoming requests to infer information about the user.

\subsection{Threat\#1: Network-Side Address Linkability}~\label{sub:network-threat1}

\textbf{Adversary}: An external service endpoint that receives address-bearing RPC requests from the wallet.\\
\textbf{Source of Leakage}: Multiple wallet-initiated calls whose payloads or timing reveal structural relationships between addresses.


Browser-extension wallets routinely automatically initiate background RPC requests (for balances, nonces, etc.) which include one or more wallet addresses in the JSON-RPC payload. Because an external service endpoint is the TLS termination point, it receives these requests in plaintext and can observe the addresses, the ordering of requests, and sometimes contact analytics and send sensitive information.

We consider the privacy threat that arises when such requests reveal that multiple blockchain addresses belong to the same wallet instance. Two observable patterns enable this inference: 

\begin{itemize}
    \item \textbf{Address Co-Occurrence}: Two or more addresses appear together in the same request (e.g. batched balance queries). This provides a \textbf{definitive} linkability signal that the addresses originate from the same wallet instance.
    \item \textbf{Address Timing Correlation}: Requests containing different addresses are sent within a short time window. This provides only a \textbf{heuristic} signal: closely timed lookups may suggest that the addresses are managed by the same wallet instance, but this inference is less reliable and becomes more informative when combined with other metadata (e.g., IP address).
\end{itemize}

An external endpoint observing either of these patterns can infer structural relationships between a user’s addresses, even without user interaction or webpage involvement.

\subsection{Network-Side Results}

We applied the improved request-interceptor framework to all three wallet datasets and analyzed the address-bearing traffic emitted during measurement procedure. Our analysis quantifies (i) the prevalence of address exposure to external endpoints and (ii) the extent to which these requests reveal structural relationships between multiple addresses within the same wallet.

\begin{table}[t!]
\centering
\small

\begin{tabular}{p{4.5cm}x{0.9cm}x{0.9cm}x{0.9cm}}
\toprule
\textbf{Metric} &
\textbf{\TorresOld{}} &
\textbf{\TorresNew{}} &
\textbf{\NewEightyFive{}} \\

\midrule
\multicolumn{4}{l}{\textbf{Number of Wallets Leak Addresses}} \\
\midrule
To any endpoints & 42 & 43 & 57 \\
To analytics endpoints & 0 & 2 & 3 \\
To telemetry endpoints & 4 & 4 & 4 \\

\midrule
\multicolumn{4}{l}{\textbf{Third-Party Domains Receiving Addresses}} \\
\midrule
Total domains receiving addresses & 64 & 80 & 113 \\  
Contacted by only one wallet & 53 & 63 & 99 \\
Percentage contacted by only one wallet &
82.8\% (53/64) &  78.8\% (63/80) &  87.6\% (99/113)\\

\midrule
\multicolumn{4}{l}{\textbf{Analytics / Telemetry Presence in Wallets}} \\
\midrule
Wallets embedding analytics/telemetry &
14/42 (33.3\%) & 16/43 (37.2\%) & 26/57 (45.6\%) \\
\midrule
\end{tabular}

\begin{tabular}{lx{2.2cm}x{2.2cm}x{2.3cm}}
\multicolumn{4}{l}{\textbf{Third-Party Domains Receiving Addresses (most contacted)}} \\

\midrule
1st &
etherscan.io (9) & etherscan.io (7) & infura.io (5) \\
2nd &
infura.io (9) & infura.io (6) & aptoslabs.com (4) \\
3rd &
binance.org (4) & avax.network (5) & sentry.io (3) \\
4th &
sentry.io (3) & binance.org (4) & publicnode.com (3) \\
5th &
rabby.io (2) & sentry.io (3) & arbitrum.io (3) \\

\bottomrule
\end{tabular}

\vspace{1em}
\caption{Comparison of wallet address leakage and third-party connectivity across the \TorresOld{}, \TorresNew{}, and \NewEightyFive{} datasets.}
\label{tab:3-datasets-comparison}
\vspace{-4em}
\end{table}

\subsubsection{\textbf{Improved Coverage of Address Exposure in Legacy Datasets}}

We capture address-bearing traffic from \TorresOld{} using our improved network interceptor framework. Torres et al. reported address leakage only to non-vendor-operated domains, filtering out vendor-operated endpoints using a hand-curated list of “valid third-party” domains.\footnote{Hand-curated domain list used by Torres et al. to filter vendor-operated endpoints in their public analysis code: \url{https://github.com/christoftorres/Web3-Privacy/blob/main/wallet-address-leakage/analysis/find-leaks-and-scripts-wallet-extensions.py\#L228}} To enable a fair comparison, we apply the same filtering and then measure the additional non-vendor exposure captured by our instrumentation.

In \TorresOld{}, 12 wallets leaked addresses to non-vendor domains from the \ttpage{} context. Seven of these overlap with the 13 wallets identified in the original study. When extension background contexts \backgroundPage{} and \serviceWorker{} are included, an additional 25 wallets leaked addresses to non-vendor domains, representing a substantial increase in observable third-party exposure.

To assess whether this measurement gap still exists two years later, we apply the same procedure to \TorresNew{}. After vendor filtering, 18 wallets leaked addresses to non-vendor domains from the \ttpage{} context, while 22 did so from background contexts. Across both datasets, background activity missed by the baseline framework remains a major channel through which wallets transmit address-bearing traffic to external endpoints.

\subsubsection{\textbf{Cross-Dataset Summary of Third-Party Connectivity}}

We applied the improved framework across \TorresOld{}, \TorresNew{}, and \NewEightyFive{}. While the first two datasets enable a time-separated comparison on the same set of extensions, \NewEightyFive{} provides a broader and more recent snapshot of today’s browser-wallet ecosystem. After excluding broken and some partially functional extensions, 42, 43, and 57 wallets respectively produced analyzable address-bearing traffic.

We no longer attempt to filter vendor-operated domains from other third-party domains, since both are external to the user. Table~\ref{tab:3-datasets-comparison} categorizes the wallet address leakage and third-party connectivity across these three datasets.

Across all datasets, address leakage to RPC endpoints is common, while transmissions to analytics or telemetry services are comparatively rare. The modern ecosystem is highly fragmented: in \NewEightyFive{}, nearly 90\% of domains receiving address-bearing requests are contacted by only a single wallet. This indicates a shift away from shared, easily identifiable RPC providers (like {\small\texttt{infura.io}} and {\small\texttt{etherscan.io}}) toward wallet-specific backend domains. Many wallets proxy their RPC requests through proprietary vendor-operated domains, which obscures the identity of the underlying node provider while centralizing visibility of users’ address queries within the wallet vendor itself.

\begin{table}[ht!]
\small
\begin{tabular}{%
p{1.4cm}
>{\centering\arraybackslash}p{0.13cm}
@{\hspace{8pt}}>{\arraybackslash}p{0.13cm}
@{\hspace{8pt}}>{\arraybackslash}p{0.13cm}
@{\hspace{8pt}}>{\arraybackslash}p{0.15cm}
>{\arraybackslash}p{1.4cm}
>{\centering\arraybackslash}p{0.2cm}
>{\arraybackslash}p{1.6cm}
c}
\textbf{Wallet} &
\adjustbox{angle=45}{\textbf{\textit{Analytics embedded}}} &
\adjustbox{angle=45}{\textbf{\textit{\# Addr. Recipients}}} &
\adjustbox{angle=45}{\textbf{\textit{Addr. to Analytics}}} &
\adjustbox{angle=45}{\textbf{\textit{Co-occurence}}} &
\adjustbox{angle=45}{\textbf{\textit{Co-occur. Dom.}}} &
\adjustbox{angle=45}{\textbf{\textit{Timing Correlation}}} &
\adjustbox{angle=45}{\textbf{\textit{Timing Dom.}}} &
\textbf{Users} \\

\midrule
MetaMask & \absentmark & 3 & \absentmark & \presentmark & metamask.io & \absentmark & & 14M \\
Phantom & \absentmark & 1 & \absentmark & \presentmark & phantom.app & \absentmark &  & 5M \\
OKX  & \presentmark & 1 & \absentmark & \absentmark & -- & \absentmark & -- & 2M \\
Ronin  & \presentmark & 3 & \absentmark & \absentmark & -- & \absentmark & -- & 1M \\
Coinbase& \presentmark & 2 & \absentmark & \presentmark & coinbase.com & \presentmark & cbhq.net & 1M \\
& & & & & & & coinbase.com & \\
Trust  & \presentmark & 1 & \absentmark & \absentmark & -- & \absentmark & -- & 1M \\
Keplr & \presentmark & 6 & \absentmark & \absentmark & -- & \absentmark & -- & 1M \\
Rabby  & \presentmark & 1 & \absentmark & \absentmark & -- & \absentmark & -- & 900K \\
Solflare& \absentmark & 1 & \absentmark & \presentmark & solflare.com & \presentmark & solflare.com & 800K \\
Backpack & \presentmark & 1 & \absentmark & \absentmark & -- & \presentmark & xnfts.dev & 600K \\
TronLink & \presentmark & 2 & \presentmark & \presentmark & tronlink.org & \presentmark & tronlink.org & 600K \\
Bitget  & \absentmark & 1 & \absentmark & \absentmark & -- & \absentmark & -- & 400K \\
Petra Aptos  & \presentmark & 3 & \absentmark & \absentmark & -- & \absentmark & -- & 400K \\
Station  & \presentmark & 2 & \presentmark & \absentmark & -- & \absentmark & -- & 300K \\
Ready  & \presentmark & 1 & \absentmark & \absentmark & -- & \absentmark & -- & 300K \\
Xverse & \absentmark & 1 & \absentmark & \absentmark & -- & \absentmark & -- & 300K \\
Martian & \presentmark & 3 & \presentmark & \presentmark & mixpanel.com & \presentmark & aptoslabs.com & 200K \\
& & & & & & & mixpanel.com & \\
Leap  & \presentmark & 4 & \absentmark & \absentmark & -- & \absentmark & -- & 200K \\
SubWallet & \absentmark & 15 & \absentmark & \absentmark & -- & \presentmark & alchemy.com& 200K \\
& & & & & & & avail.so& \\
& & & & & & & blockscout.com & \\
& & & & & & & publicnode.com & \\
& & & & & & & subscan.io & \\
& & & & & & & zora.energy & \\
Suiet & \absentmark & 1 & \absentmark & \absentmark & -- & \absentmark & -- & 200K \\
\midrule
\textbf{Total \presentmark} &
13 &  & 3 &
6 &  &
6 &  &
\\
\bottomrule
\end{tabular}

\raggedright\footnotesize
\vspace{1em}
\emph{\presentmark\ indicates the presence of the corresponding behavior, while \absentmark\ indicates its absence.}
\vspace{1em}
\caption{Linkability indicators for the 20 most installed wallets in \NewEightyFive{} dataset, covering analytics use, third-party recipients of wallet addresses, address leaks to analytics sites, co-occurrence and timing signals, and user counts.}
\label{tab:5.1-linkability}

\vspace{-3em}
\end{table}

\subsubsection{\textbf{Address Correlation in Multi-Account Wallets (\NewEightyFive{})}}

We next evaluated whether outbound RPC traffic reveals relationships between multiple accounts within the same wallet. Among the 57 analyzable wallets in \NewEightyFive{}, 55 allowed the creation of at least three working test accounts.

We consider two correlation indicators defined in Section~\ref{sub:network-threat1}: (i) multiple addresses \textit{co-occur} within the same request, and (ii) requests containing different addresses showing \textit{timing correlation}. Across the 55 multi-account wallets, 13 leaked two or more test addresses in the same request to at least one endpoint, and 11 exhibited timing correlation (within 10ms time window) between requests containing different addresses. In total, \textbf{17 wallets exposed at least one of these correlation signals}. An estimated 22.1 million users (\textbf{62.9\%}) of 35.16 million users in \NewEightyFive{} are affected by co-occurrence and 23.0 million users (\textbf{65.4\%}) are affected by at least one of the two signals. 

Table~\ref{tab:5.1-linkability} shows these behaviors for the twenty most widely installed wallets in \NewEightyFive{}. The table reports whether any analytics site is embedded, the number of distinct domains that receive address-bearing requests, whether any of these domains are analytics services, and the observed linkability signals, including address co-occurrence, timing correlation, and the specific domains to which these signals were leaked.

\subsection{Discussion}

These results show that network-side privacy exposures stem directly from routine wallet behavior. Because wallets automatically issue address-bearing RPC requests in the background, external endpoints gain visibility into users’ addresses and, for many wallets, the relationships between multiple addresses. In particular, address co-occurrence directly reveals that multiple addresses originate from the same wallet instance, while timing correlation provides a weaker but still informative indicator of such relationships.

We also observe a shift toward wallet-specific backend domains. This centralizes visibility within individual vendors while increasing the number of parties that receive address-bearing traffic.

In practice, users have little control over which entities learn their network activity. These findings show that users implicitly entrust wallet vendors and the external services the vendors rely on with sensitive address-level information. For actively malicious (or compromised) vendors, such visibility could enable targeted spear-phishing campaigns against high-value users~\cite{extropyPhishing2025,guan2024characterizing}.

\section{Web-Side Threats and Results} \label{sec:web-side}

\subsection{Threat Model}\label{subsub:threat-model:web-adversary}

The adversary is any third-party tracker script intentionally included by an otherwise benign webpage, for example for analytics, performance monitoring, telemetry, or user-experience features. Once loaded, such scripts execute with the same origin privileges as the embedding webpage and can access wallet provider objects exposed to the webpage via EIP-6963. As a result, they can observe:

\begin{itemize}
    \item the presence of installed wallets through EIP-6963 discovery events, for example, \announceProvider{} and \requestProvider{}
    \item the return value of the restricted method \ethAccounts{}, which may silently reveal addresses previously authorized for the webpage origin, since third-party scripts embedded in the page execute under that same origin
\end{itemize}

The adversary executes arbitrary JavaScript in accordance with browser semantics. It cannot bypass the Same-Origin Policy (SOP).

The adversary can access all currently available wallet provider methods. Restricted methods such as \ethAccounts{} only yield addresses if the user previously granted permission to the webpage origin and that permission has not been revoked. The adversary passively inspects these responses but does not alter wallet behavior. It cannot approve wallet prompts, and cannot extract private keys or authorize restricted operations without explicit user interaction. 

We also do not assume access to persistent browser storage (e.g., cookies or localStorage), representing users who routinely clear such state or employ privacy-preserving browser settings. 

We assume the dApp tries to revoke the permission to the wallet from their side properly by using the \revokePermissions{} method.

\subsection{Threat\#1: Wallet-Based Fingerprinting via EIP-6963 Discovery}

\textbf{Adversary}: A third-party tracker script embedded on any websites (not limited to dApps).\\
\textbf{Source of Leakage}: EIP-6963 provider discovery events.
\smallskip

EIP-6963 enables websites to request a response from all the installed wallets by dispatching the \requestProvider{} event on the window. Each EVM-compatible wallet responds with an \announceProvider{} event containing its name and provider metadata. This allows a third-party tracker script on any website to learn: 

\begin{itemize}
    \item which wallets the user has installed,
    \item how many wallets are present,
    \item the exact combinations of wallets, which forms a \textbf{stable and potentially distinguishing} fingerprint.
\end{itemize}

Wallet-installation patterns are more distinctive than common browser fingerprints (fonts, canvas, etc.) and typically change slowly. Industry reports also indicate that many cryptocurrency users manage more than one wallet~\cite{reownSurvey2025,coinlawsurvey2026}, suggesting a greater diversity of possible installed-wallet combinations. Because discovery requires no user interaction, this fingerprinting attack applies to:

\begin{itemize}
    \item users who never connected a wallet to the site,
    \item users who block cookies or clear browser storage. 
\end{itemize}

As a result, EIP-6963 provides a new browser-level fingerprinting vector targeting at the Web3 wallet users.

\subsection{Threat\#2: Cross-Session Tracking and Addresses Clustering}\label{subsec:threat-2}

\textbf{Adversary}: A third-party tracker script embedded on a dApp\\
\textbf{Source of Leakage}: The lack of a standardized dApp-side permission revocation mechanism. 
\vspace{0.5em}

EIP-2255 does \textit{not} define any expiration, renewal, or time-based invalidation of permissions, and there is no revocation standard. Trackers embedded on a dApp can see the granted addresses of the dApp by calling \ethAccounts{} without user interaction provided that the wallet has not revoked the dApp’s permission to access its addresses.

We call a wallet \textbf{revocation-unsafe} if the wallet continues returning previously authorized addresses through \ethAccounts{} after receiving a revocation request \revokePermissions{} from a dApp.




\begin{figure}[t!]
    \centering
    \includegraphics[width=0.75\columnwidth]{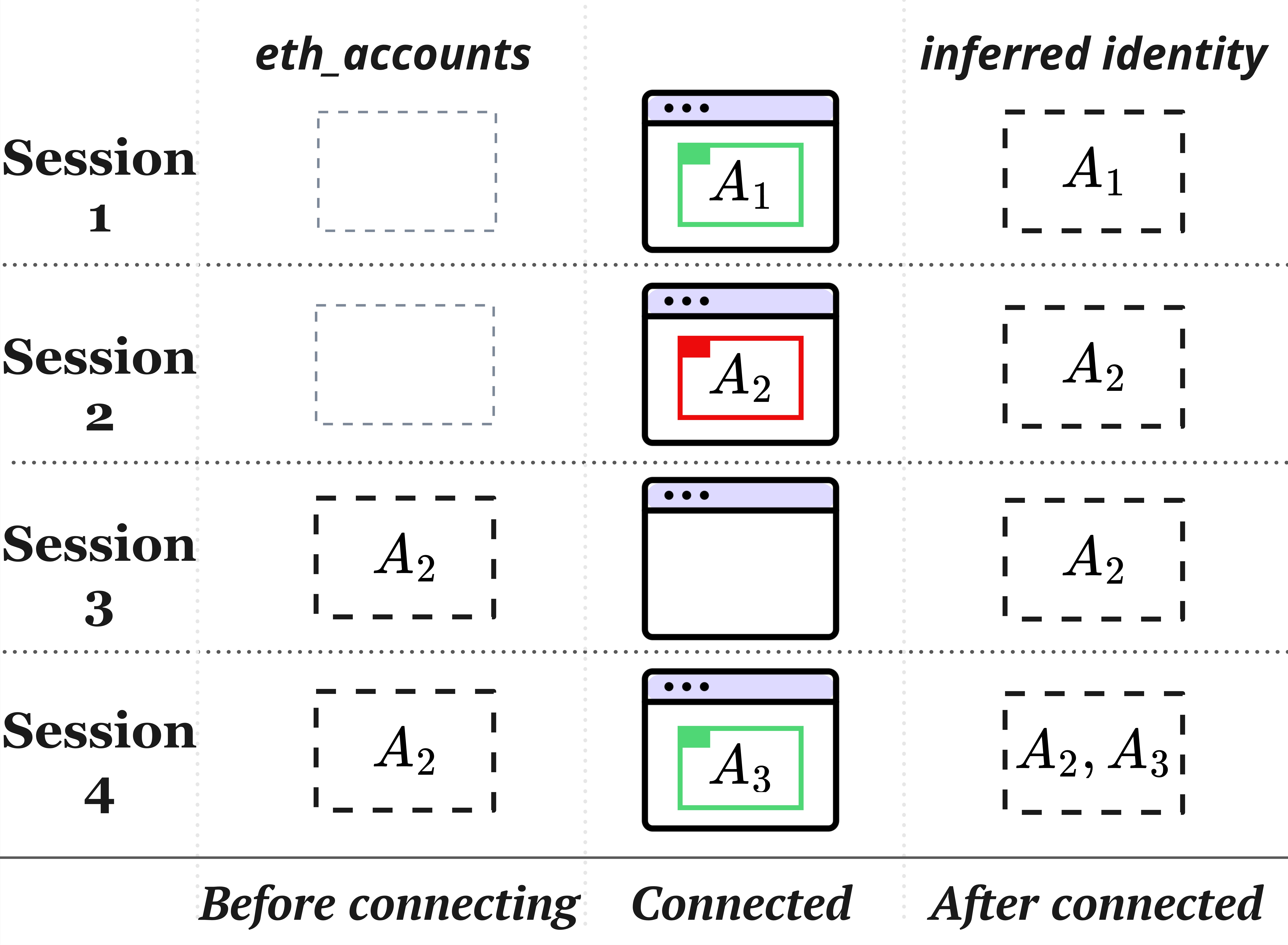}
    \caption{Single-site cross-session re-identification and address clustering via stale addresses returned by revocation-unsafe wallets.}
    \label{fig:threat-2}
\end{figure}

Figure~\ref{fig:threat-2} shows cross-session re-identification and addresses clustering. Suppose the user has at least one revocation-unsafe wallet: 

\begin{enumerate}
\item \textbf{Session 1}: The user connects with address $A_1$.
The wallet $W_p$ correctly revokes this permission when the user disconnects via the dApp UI (green border).
\item \textbf{Session 2}: The user revisits the dApp, connects a different wallet $W_q$ and grants $A_2$, but $W_q$ does not remove the site's permission when the dApp requests a disconnect (red border).
\item \textbf{Session 3}: The user revisits the dApp without any action. The tracker still receives $A_2$ from $W_q$ via \ethAccounts{}.
\item \textbf{Session 4}: The user revisits the dApp, connects wallet $W_p$ and grants a fresh address $A_3$. The stale $A_2$ continues to appear.
\end{enumerate}

The tracker can therefore conclude: 

\begin{itemize}
    \item The same user visited in sessions 2, 3, and 4,
    \item Addresses $A_2$ and $A_3$ belong to the same user.
\end{itemize}



Because blockchain addresses are globally unique and cannot be produced in a wallet without owning the private key, \textbf{the stale entry functions as a very strong and more durable identifier}, even if the user rotates wallets or switches accounts. Unlike cookies or localStorage values, $A_2$ is stored in extension-controlled storage and cannot be cleared, overwritten, or spoofed. The only mitigation available now is for the user to manually clean up the permissions inside the wallet's ``Connected Sites'' menu.

Hence, these stale entries persist across \textbf{page reloads}, \textbf{browser restarts}, and even after the user \textbf{clear their browser cache}.




\subsection{Threat\#3: Cross-Site Tracking and Addresses Clustering across dApps} \label{subsec:threat-3}

\textbf{Adversary}: A shared third-party tracker script embedded on multiple dApps.\\
\textbf{Source of Leakage}: The lack of a standardized dApp-side permission revocation mechanism.


Cross-site tracking arises when multiple dApps share the same third-party tracker. With first-party privileges on each site, the script can detect all the installed wallets, and invoke \ethAccounts{} independently on every dApp where it appears.


\begin{figure}[t!]
    \centering
    \includegraphics[width=0.73\columnwidth]{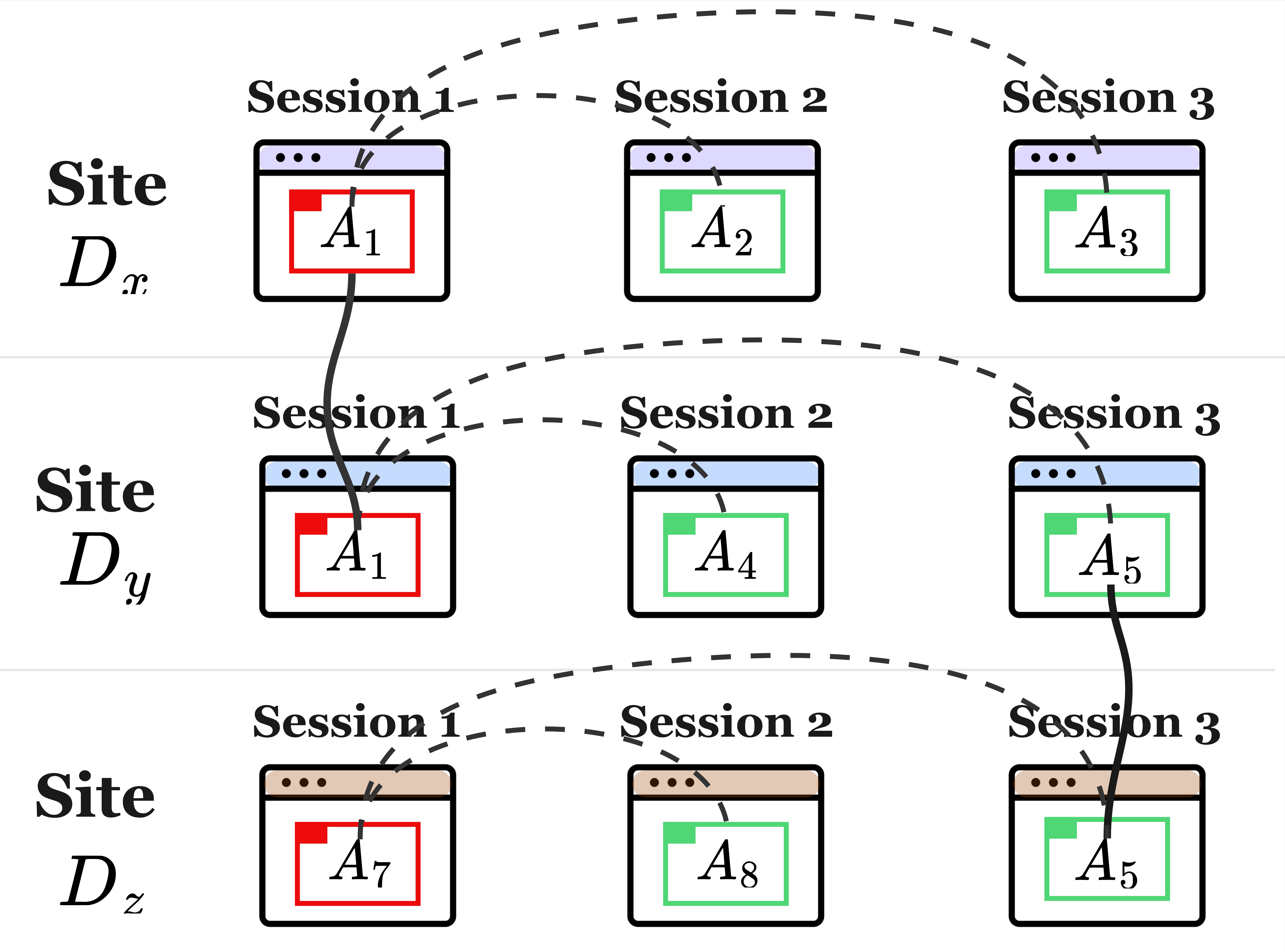}
    \caption{Cross-site cross-session tracking and wallet-address clustering enabled by stale permissions and shared third-party trackers.}
    \label{fig:threat-3}
\end{figure}

If the user has at least one \textit{revocation-unsafe} wallet, the stale address returned by that wallet becomes visible to the shared tracker on that site. Figure~\ref{fig:threat-3} illustrates how this results in cross-site tracking and multi-address clustering.

\subsubsection{\textbf{Tracking}}
Suppose the user previously granted address $A_1$ to dApp $D_x$ using a revocation-unsafe wallet. Even if the user later switches wallets or accounts, the stale entry $A_1$ continues to appear through \ethAccounts{}, allowing sessions 2 and 3 on $D_x$ to be linked to the same user. The same occurs independently on dApp $D_y$.

If a third-party tracker $G$ is embedded in both $D_x$ and $D_y$: on Site $D_x$, $G$ sees the stale address $A_1$, and on Site $D_y$, $G$ also sees $A_1$. 
Therefore, the tracker can immediately infer that the user visiting $D_x$ and $D_y$ is the same individual.

\subsubsection{\textbf{Wallet Clustering}}

As we discussed in Threat~\ref{subsec:threat-2}, all subsequent sessions on $D_x$ and $D_y$ remain linkable. On $D_x$, the tracker observes addresses $\{A_1, A_2, A_3\}$ across sessions. On $D_y$, the tracker separately observes $\{A_1, A_4, A_5\}$. Because both sets contain the same stale identifier $A_1$, the tracker can merge them into a single cluster: $\{A_1, A_2, A_3, A_4, A_5\}$.

A similar process occurs on Site $D_z$, where a different stale identifier $A_7$ creates a separate cluster $\{A_7, A_8\}$. These two clusters remain disjoint until any address appears in both. In the example of Figure~\ref{fig:threat-3}, the appearance of $A_5$ on $D_z$ links $D_y$ and $D_z$, enabling the tracker to conclude: 

\begin{itemize}
    \item Sites $D_x$, $D_y$, and $D_z$ were visited by the same user. 
    \item The combined address set $\{A_1, A_2, A_3, A_4, A_5, A_7, A_8\}$ belong to that user. 
\end{itemize}

This cross-site clustering extends the leakage beyond a single origin. While Threat\#2 in Section~\ref{subsec:threat-2} links sessions within one dApp, a shared tracker aggregates stale identifiers across every dApp that embeds it. As a result, \textbf{a single revocation-unsafe wallet enables third-party domains to build ecosystem-wide profiles}, merging wallet addresses and activity patterns observed on multiple unrelated sites. The more widely a tracker is embedded, the larger the portion of the user's Web3 activity it can reconstruct.








\subsection{Threat\#4: Tracking and Deanonymization beyond dApps} \label{subsec:threat-4}

\textbf{Adversary}: A third-party tracker script shared between a website and a previously connected dApp.\\
\textbf{Source of Leakage}: The combination of (i) wallet provider exposure in cross-origin iframes, (ii) dApps that permit iframe embedding.
\vspace{0.5em}

Whereas Threat\#3 requires that a user at least once grant the same address to different dApps for a tracker to link two address clusters (see Figure~\ref{fig:threat-3}, Sites $D_y$ and $D_z$), we now present a more general attack that does not require this assumption. This attack is also applicable to any website, not just dApps.

Assume that address $A_1$ has previously been granted by the wallet to \dappX{}. A tracker running on \websiteA{} can not access this information, because the tracker $T$ is confined to the origin of the site it runs on, and wallets strictly validate the origin of any incoming request.

\begin{figure}[t!]
    \centering
    \includegraphics[width=0.8\columnwidth]{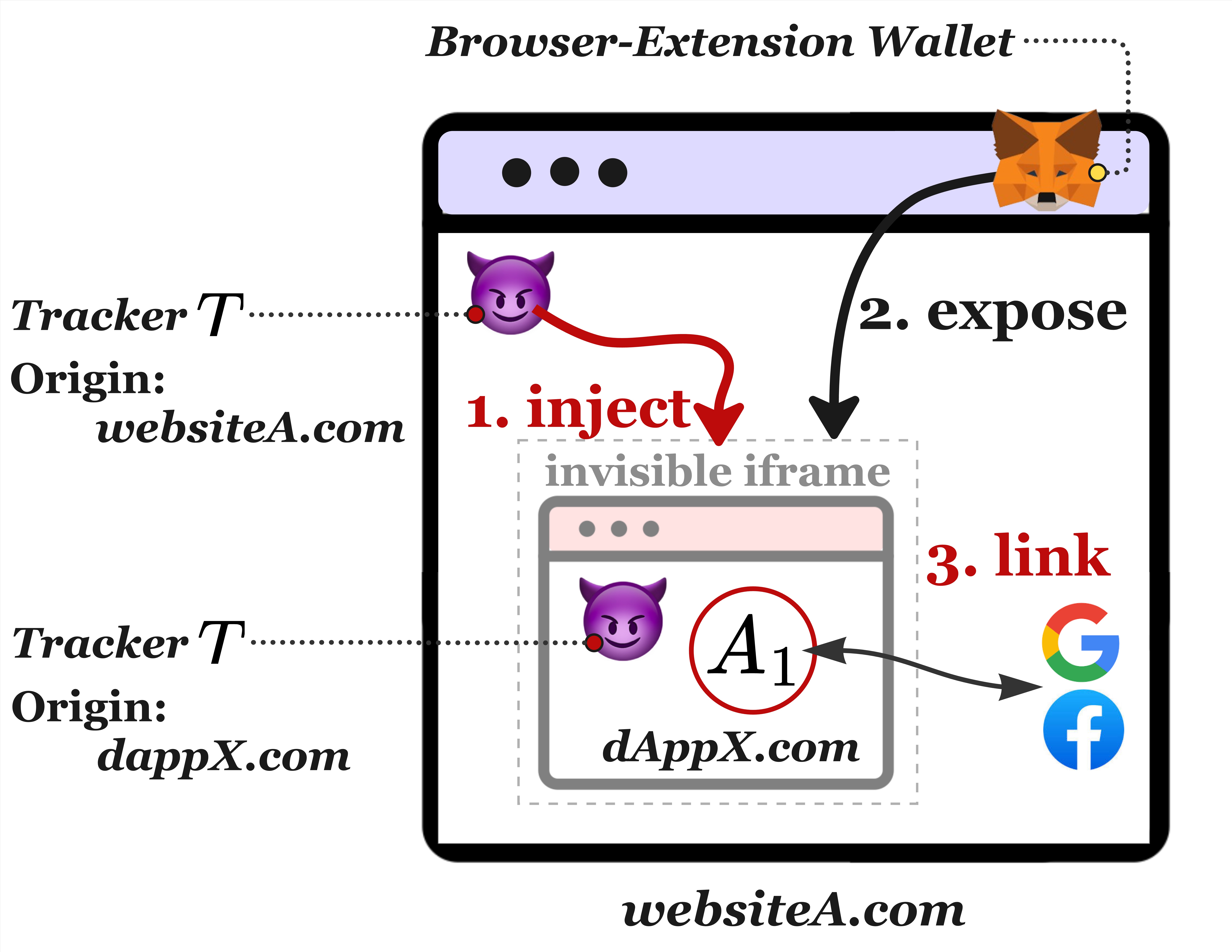}
    \caption{A shared third-party tracker on \websiteA{} embeds an invisible iframe of \dappX{}. If the wallet exposes its provider to cross-origin iframes, the tracker inside the iframe can read the user's previously authorized address and pass it back to the tracker on \websiteA{}.}

    \label{fig:threat-4}
    \vspace{-1em}
\end{figure}

However, as seen in Figure~\ref{fig:threat-4}, tracker $T$ on \websiteA{} can embed one (or more) invisible iframes in order to load iframe-embeddable dApps which also include $T$. 

For wallets that expose their provider objects to cross-origin iframes and do not restrict provider access to the top-level browsing context, the instance of $T$ executing inside such an iframe runs as a same-origin script on the embedded dApp's origin and can invoke \ethAccounts{}. If the user has previously granted wallet access to that dApp, the address becomes visible to the tracker inside the iframe.

The observed address can then be passed from the tracker $T$ inside the iframe to the tracker $T$ on \websiteA{}, allowing it to identify the user and link the address to the user's activity on that site. 

The tracker does not need to know in advance which dApp, if any, the user has previously connected. It can embed one or more candidate dApps that also include the same tracker and test them opportunistically. The attack succeeds for any embedded dApp to which the user previously granted wallet access and for which the wallet exposes its provider inside the iframe.

\subsubsection{\textbf{Cross-Session Tracking on Any Website}}

Once extracted, the information of the address can be used as a stable, cross-session identifier on any website. Because wallet addresses are globally unique, and the permission persists if the user never revokes it, the tracker gains a unique and long-lived identifier for the user, enabling cross-session tracking on that website. 

\subsubsection{\textbf{Linking Web Activities to Web3 Identities and On-Chain Wealth}}

The exposed wallet address serves as a persistent identifier across different websites that embed the same tracker. As a result, the tracker can link a user's activity across sites. Combined with the address clustering described in Threat~\ref{subsec:threat-3}, this allows the tracker to associate diverse Web2 activities, including news consumption, shopping behavior, and search queries, with the same cluster of Web3 addresses revealing the user's on-chain wealth.

\subsubsection{\textbf{Deanonymization of Web3 Identities}}

If the website exposes any real-world user attributes in the DOM (e.g., email address, display name, phone number, or identifiers associated with Google or Facebook logins), these become visible to the tracker's JavaScript. The tracker can therefore link: 

\begin{itemize}
    \item \textbf{Web2 identity}: user information exposed by the site,
    \item \textbf{Web3 identity}: the user's wallet address obtained by the injected iframe, and the address clusters from Threat~\ref{subsec:threat-3}.
\end{itemize}

A single popular dApp can be sufficient for this attack. Once the tracker is able to extract a wallet address through an iframe-embeddable dApp for which the user has previously granted access, any website embedding the same tracker can contribute additional Web2 or Web3 identity signals. As the tracker appears across more websites, it can reconstruct an increasingly complete view of the user's combined Web2 and Web3 activity.

\subsection{Web-Side Results}

\subsubsection{\textbf{EIP-6963 Discovery Support}}

Of the 85 wallets in \NewEightyFive, 36 exposed an EVM-compatible provider interface and could be detected without unlocking by EIP-6963 events.
Importantly, despite representing only ~42\% of the wallets in the dataset, these 36 EVM-compatible wallets collectively have 29.08 million Chrome Web Store users, \textbf{which is 82\% of the total 35.16 million users represented in \NewEightyFive{}}. This indicates that our selection captures the majority of the user base in the dataset.


\subsubsection{\textbf{Wallet Permission Revocation Behavior}}

We evaluated how wallets in \NewEightyFive{} handle permission revocation and whether a dApp can still access previously granted addresses after a disconnect action. 


\paragraph{\textbf{Analysis Scope}}

These 36 wallets all exposed an EIP-6963 discovery interface, allowing programmatic interaction through standard provider APIs. They constitute all EVM-compatible wallets in the dataset; non-EVM wallets did not expose a compatible provider and were excluded from further analysis.

\begin{table}[t]
  \centering
  \small
  \caption{Revocation behavior among the 20 most widely used EVM-compatible browser-extension wallets ($\geq$ 100{,}000 Chrome Web Store users).}
  \label{tab:wallet-revoke-top20}
  \begin{tabular}{lccc}
    \toprule
    Wallet & Revocation unsafe & Error returned? & Users \\
    \midrule
    MetaMask & \absentmark & -- & 14M \\
    Phantom & \presentmark & \textit{Not supported} & 5M \\
    OKX Wallet & \presentmark & -- & 2M \\
    Ronin Wallet & \presentmark & \textit{Not supported} & 1M \\
    Coinbase Wallet & \presentmark & \textit{Not supported} & 1M \\
    Trust Wallet & \presentmark & -- & 1M \\
    Keplr & \absentmark & -- & 1M \\
    Rabby Wallet & \absentmark & -- & 900K \\
    Backpack & \presentmark & \textit{Not supported} & 600K \\
    Bitget Wallet & \presentmark & \textit{Not supported} & 400K \\
    Ctrl Wallet & \absentmark & -- & 300K \\
    Bybit Wallet & \presentmark & \textit{Not supported} & 200K \\
    Leap Wallet & \absentmark & -- & 200K \\
    SubWallet & \absentmark & -- & 200K \\
    Gate Wallet & \presentmark & \textit{Not supported} & 100K \\
    Zerion Wallet & \presentmark & \textit{Not supported} & 100K \\
    Pontem & \presentmark & \textit{Not supported} & 100K \\
    Coin98 & \presentmark & -- & 100K \\
    Portal DEX & \presentmark & -- & 100K \\
    Exodus Web3 Wallet & \absentmark & -- & 100K \\
    \bottomrule
  \end{tabular}

  \par\vspace{10pt} 
  \raggedright\footnotesize
  
  \emph{``Revocation unsafe'' indicates whether invoking \texttt{wallet\_revokePermissions}
  removed the origin's permission entry from the wallet's internal state
  (\presentmark\ = permission persisted; \absentmark\ = permission successfully cleared,
  such that subsequent \texttt{eth\_accounts} calls returned no addresses})

  \smallskip

  \emph{``Error returned?'' indicates whether the wallet returned an explicit error in
  response to the revocation request (e.g., ``method not supported''). ``--'' denotes
  that the wallet returned no error message, regardless of whether revocation
  succeeded.}

  \vspace{-1em}
\end{table}

\paragraph{\textbf{Revocation Support}}\label{par-results-revocation-support}

Of these discoverable 36 wallets, 22 (61.1\%) did not correctly implement permission revocation. Of these, 15 explicitly returned an error indicating that the revocation method was unsupported (e.g., ``\textit{wallet\_revokePermissions does not exist/is not available}''). Such errors are surfaced to the dApp but not displayed to the user. The remaining 14 wallets successfully removed the permission entry associated with the testing origin.

Table~\ref{tab:wallet-revoke-top20} summarizes revocation behavior for the 20 most widely installed EVM-compatible wallets ($\geq$100{,}000 Chrome users). A complete list of all 36 discoverable wallets is provided later in Table~\ref{tab:wallet-iframe-lock-unsafe}.


\paragraph{\textbf{Post-Revocation Behavior}}

All 22 wallets that failed to revoke permissions continued to return the previously granted address when queried with \ethAccounts{}, even after a new session. This behavior also persisted sometimes when the wallet UI was explicitly locked. This demonstrates that these wallets retain stale permission state internally and continue to expose user addresses to the same origin despite receiving a revocation request.



\subsubsection{\textbf{dApp Permission and Tracking Behavior}}

We analyzed 30 popular Ethereum dApps from DappRadar’s Ethereum category to observe how real sites handle wallet permissions and user disconnection (see Table~\ref{tab:dapp-summary}).


\paragraph{\textbf{Invocation of Permission Revocation}}
Only 11 out of 30 dApps (36.7\%) invoked a \revokePermissions{} method when the user clicked a ``Disconnect,'' ``Logout,'' or similar UI control on the site. The remaining 19 dApps performed the disconnect action only at the application level: they cleared their own interface state but never issued a revocation request to the wallet provider.

\paragraph{\textbf{Probe Accounts Before Explicit Permission}}

18 of the 30 dApps (60\%) called the restricted method ``eth\_accounts'' before requesting any permission-granting API. Wallets are required to return an empty array when no permission has been granted, but any wallet with stale permission state will still return the previously authorized addresses. This means pre-permission probes can reveal a user’s address if the wallet does not handle revocation correctly.

\begin{table}[t]
  \centering
  \small
  \begin{tabular}{lcc}
    \toprule
    \textbf{Behavior} & \textbf{Count} & \textbf{Percentage} \\
    \midrule
    Do not invoke \revokePermissions{} & 19 & (19/30) 63.3\% \\
    Calls \ethAccounts{} before permission & 18 & (18/30) 60.0\% \\
    Stores address in localStorage/cookies & 27 & (27/30) 90.0\% \\
    Do not clear stored address on logout (of 27) & 17 & (17/27) 63.0\% \\
    Contacts $\geq$1 third-party tracker & 19 & (19/30) 63.3\% \\
    Contacts $\geq$3 third-party trackers & 14 & (14/30) 46.7\% \\
    \bottomrule
  \end{tabular}
  \vspace{1em}
  \caption{Summary of permission- and tracking-related behaviors observed across 30 popular Ethereum dApps.}
  \label{tab:dapp-summary}
  \vspace{-3em}
\end{table}

\paragraph{\textbf{Client-Side Storage of Wallet Addresses}}

Our framework monitored changes to browser storage during the workflow. 27 of the 30 dApps (90\%) stored the connected wallet address in browser-side storage, either localStorage or cookies, at some point during the session. These identifiers survive page reloads and enable re-identification within the same site. 17 of these 27 dApps did not clear the stored address upon logout.

\paragraph{\textbf{Presence of Third-Party Analytics and Trackers}} \label{par:results-dapp-trackers}

Outbound request logs show that 19 out of 30 dApps (63.3\%) contacted at least one third-party analytics or telemetry service during the workflow, and 14 of those contacted three or more distinct tracking domains. The most common destinations were: 

\begin{itemize}
    \item \textbf{Google Analytics} / \textbf{Google Tag Manager} (13 dApps)
    \item \textbf{Sentry} (8 dApps)
    \item \textbf{Intercom} (4 dApps)
\end{itemize}

These findings show that many dApps incorporate third-party analytics or telemetry services as part of their client-side operation.

\subsubsection{\textbf{Iframe Exposure of Wallets and dApps}}

Threat~\#4 (Section~\ref{subsec:threat-4}) requires both (i) a wallet that exposes its provider inside a cross-origin iframe and does not restrict provider access to the top-level browsing context, and (ii) a dApp whose frontend can be embedded as an iframe. Therefore, we measured these two dimensions for all wallets and dApps in our dataset \NewEightyFive.

\paragraph{\textbf{dApps Iframe Embeddability}}

For Threat\#4, a malicious site must embed a real dApp inside an invisible iframe. Across the 30 Ethereum dApps we analyzed, 18 were fully iframe-embeddable, including several widely used platforms such as \textit{Uniswap}, \textit{Aave}, \textit{Lido}, and \textit{PancakeSwap}. Only 12 dApps set framing protections (e.g., {\small{\texttt{X-Frame-Options: DENY}}} or CSP {\small\texttt{frame-ancestors}}) that prevent their frontend from being loaded inside a cross-origin iframe. Iframe-embeddable dApps give an attacker a usable execution environment where an embedded tracker script can access a wallet provider that exposes itself inside the iframe.

\paragraph{\textbf{Wallet Iframe Exposure and Locked-State Leakage}}

\begin{table}[t]
\centering
\small

\begin{tabular}{@{}l l@{}}  
%
\begin{tabular}[t]{l
@{\hspace{8pt}}>{\arraybackslash}p{0.13cm}
@{\hspace{8pt}}>{\arraybackslash}p{0.16cm}
@{\hspace{8pt}}>{\arraybackslash}p{0.18cm}
c}

\textbf{Wallet} &
\rotatebox{60}{\textbf{Iframe Exposure}} &
\rotatebox{60}{\textbf{Leaks When Locked}} &
\rotatebox{60}{\textbf{Revoc. Unsafe}} &
\textbf{Users} \\ \midrule

MetaMask        & \presentmark & \presentmark   & \absentmark & 14M \\
Phantom         & \presentmark & \presentmark   & \presentmark   & 5M \\
OKX             & \presentmark & \presentmark   & \presentmark   & 2M \\
Ronin           & \presentmark & \absentmark & \presentmark   & 1M \\
Keplr           & \presentmark & \absentmark & \absentmark & 1M \\
Trust           & \presentmark & \presentmark   & \presentmark   & 1M \\
Coinbase        & \presentmark & \presentmark   & \presentmark   & 1M \\
Rabby           & \presentmark & \absentmark & \absentmark & 900K \\
Backpack        & \presentmark & \presentmark   & \presentmark   & 600K \\
Bitget          & \absentmark & --        &  \presentmark     & 400K \\
Ctrl            & \absentmark & --        &  \absentmark   & 300K \\
Bybit           & \presentmark & \absentmark & \presentmark   & 200K \\
Leap            & \presentmark & \absentmark & \absentmark & 200K \\
SubWallet       & \presentmark & \presentmark   & \absentmark & 200K \\
Zerion          & \presentmark & \absentmark & \presentmark   & 100K \\
Coin98          & \presentmark & \presentmark   & \presentmark   & 100K \\
Gate            & \presentmark & \presentmark   & \presentmark   & 100K \\
Pontem          & \presentmark & \presentmark   & \presentmark   & 100K \\
Portal DEX      & \absentmark & --        &  \presentmark    & 100K \\
\midrule
\end{tabular}
&
\begin{tabular}[t]{l
@{\hspace{8pt}}>{\arraybackslash}p{0.13cm}
@{\hspace{8pt}}>{\arraybackslash}p{0.16cm}
@{\hspace{8pt}}>{\arraybackslash}p{0.18cm}
c}

\textbf{Wallet} &
\rotatebox{60}{\textbf{Iframe Exposure}} &
\rotatebox{60}{\textbf{Leaks When Locked}} &
\rotatebox{60}{\textbf{Revoc. Unsafe}} &
\textbf{Users} \\ \midrule

Exodus          & \absentmark & --        &  \absentmark   & 100K \\
StarKey         & \absentmark & --        &  \absentmark   & 100K \\
Rainbow         & \absentmark & --        &  \absentmark   & 100K \\
TokenPocket     & \absentmark & --        &  \presentmark    & 90K \\
Binance         & \presentmark & \presentmark   & \absentmark & 70K \\
Core            & \presentmark & \absentmark & \presentmark   & 60K \\
MathWallet      & \presentmark & \presentmark   & \presentmark   & 50K \\
Enkrypt         & \absentmark & --        &  \presentmark    & 50K \\
QSafe           & \absentmark & --        &  \presentmark    & 30K \\
OneKey          & \presentmark & \presentmark        &  \absentmark  & 30K \\
Wigwam          & \presentmark & \absentmark & \presentmark   & 20K \\
Fin             & \presentmark & \presentmark   & \presentmark   & 20K \\
Fluvi           & \absentmark & --        &  \absentmark   & 20K \\
Hana            & \presentmark & \absentmark & \presentmark   & 10K \\
Stargazer       & \absentmark & --        &  \presentmark    & 10K \\
Flow            & \absentmark & --        &  \absentmark   & 10K \\
Koala           & \absentmark & --        &  \absentmark   & 10K \\
\midrule
\textbf{Total \presentmark} &
\textbf{23} & \textbf{14} & \textbf{22} & \textbf{29.08M} \\
\bottomrule
\end{tabular}
\\ 
\end{tabular}

\vspace{0.6em}
\emph{\presentmark\ indicates the presence of the behavior, while \absentmark\ indicates its absence.}
\vspace{0.6em}
\caption{Per-wallet iframe exposure, locked-state leakage, and revocation behavior (split into two columns).}

\label{tab:wallet-iframe-lock-unsafe}
\vspace{-3em}
\end{table}

Table~\ref{tab:wallet-iframe-lock-unsafe} summarizes three behaviors relevant to Threat~\#4 (Section~\ref{subsec:threat-4}):

\begin{itemize}
    \item \textbf{Iframe Exposure (23/36)}: whether the wallet injects its provider into an embedded iframe of a dApp and do not restrict access. This determines whether an attacker can access the wallet provider from within the dApp iframe.
    \item \textbf{Leaks When Locked (14/36)}: if calling \ethAccounts{} from within that iframe returns an address even when the wallet is locked. This affects whether cross-browser-session identification is possible when the wallet auto-locks when the browser is closed.
    \item \textbf{Revocation Unsafe (22/36)}: whether the wallet continues to expose previously granted addresses after a dApp requests permission revocation, which is derived from the revocation experiment.
\end{itemize}

\begin{table*}[t!]
\centering
\small
\begin{tabular}{p{5.3cm} p{7.5cm} l l}
\toprule
\textbf{Threat} & \textbf{Cause} & \textbf{\# Wallets} & \textbf{Estimated Users} \\
\midrule

\multicolumn{4}{l}{\textit{Network-Side Threat (Adversary: external service endpoints)}} \\

\#1 Address linkability 
& Address batching or short-interval network requests
& 17 / 85 (20\%) 
& $\sim$23.0M (65.4\%) \\

\midrule
\multicolumn{4}{l}{\textit{Web-Side Threats (Adversary: third-party trackers)}} \\

\#1 Wallet-based fingerprinting 
& EIP-6963 wallet discovery reveals installed wallet combinations 
& 36 / 85 (42.4\%) 
& $\sim$29.08M (82\%) \\

\#2 Cross-session tracking \& address clustering
& Stale address persistence after \revokePermissions{} call
& 22 / 36 (61.1\%) 
& $\sim$12.04M (34.2\%) \\

\#3 Cross-dApp tracking \& address clustering
& Stale address persistence after \revokePermissions{} call
& 22 / 36 (61.1\%) 
& $\sim$12.04M (34.2\%) \\

\#4 Cross-any-site tracking 
& Wallet provider exposure on cross-origin contexts
& 23 / 36 (63.9\%) 
& $\sim$27.76M (78.95\%) \\

\bottomrule
\end{tabular}
\par\vspace{7pt} 

\caption{Summary of the five privacy threats identified in browser-extension wallets, their causes, affected wallet numbers and ecosystem impact. Wallet counts and user counts are based on the \NewEightyFive{} dataset. For web-side threats \#2–\#4, percentages of affected wallets are computed over the 36 wallets responding to EIP-6963 events in web-side threat \#1, while user counts are computed over the full dataset.}
\par\vspace{-20pt} 
\label{tab:threat-summary}
\end{table*}

\subsubsection{\textbf{Feasibility and Impact of Threat~\#4}}

The measurements show that the components required for iframe-based cross-site wallet tracking are widely present in the current ecosystem. On the dApp side, 18 of the 30 Ethereum dApps we analyzed are iframe-embeddable, meaning a malicious site can load them invisibly in a cross-origin iframe. Among these 18, 11 perform no permission-revocation at all, so any address the user previously granted to these dApps remains readable to scripts executing inside the iframe no matter if user clicks ``disconnect'' or not.

\begin{figure}[h!]
    \centering
    \includegraphics[width=0.75\columnwidth]{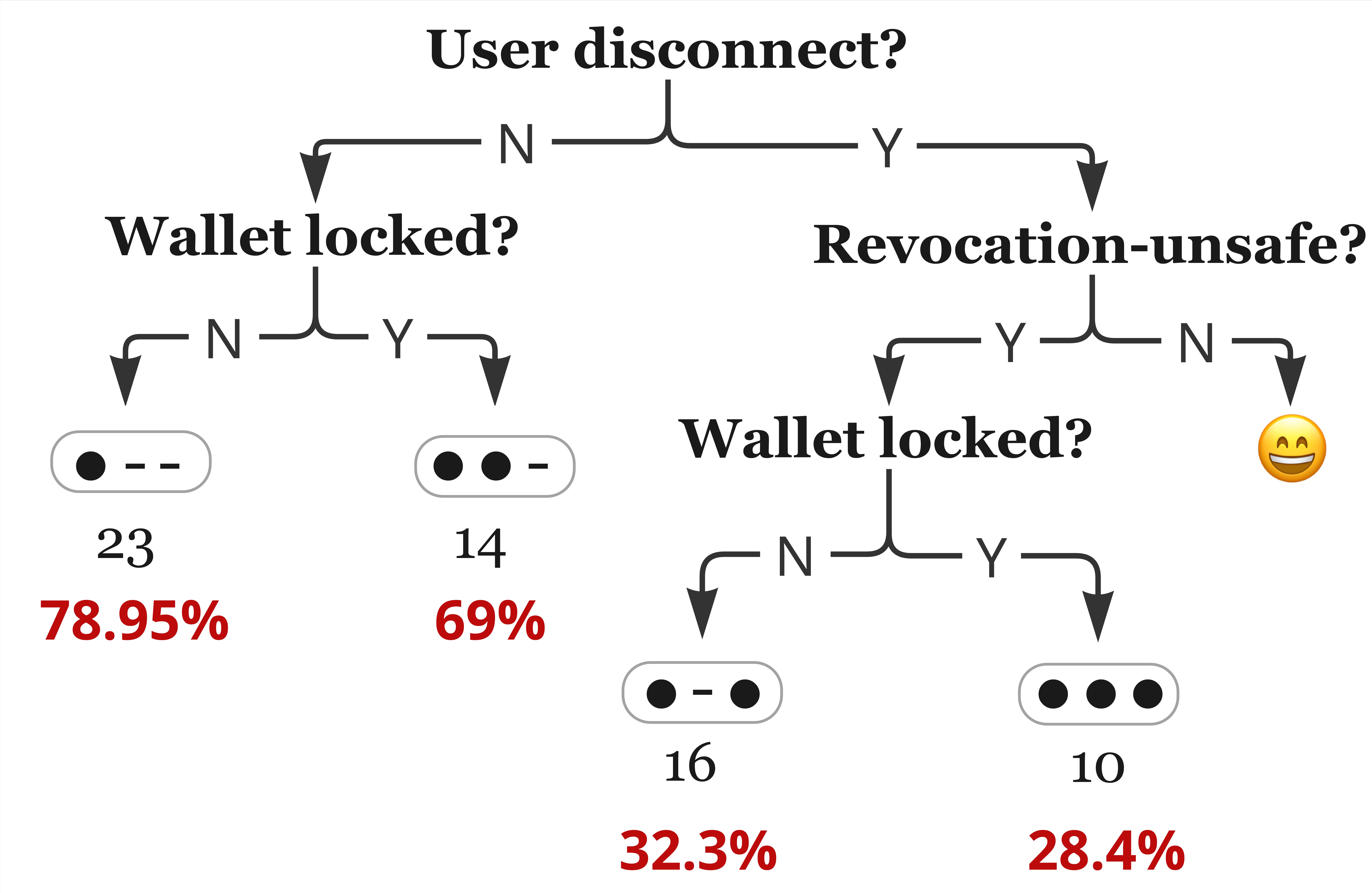}
    \caption{Outcome tree showing which wallets (referenced in Table~\ref{tab:wallet-iframe-lock-unsafe}) and what share of \NewEightyFive{} users remain vulnerable under different disconnect, revocation, and locked-state conditions.}
    \label{fig:user-decision}
    \vspace{-1em}
\end{figure}

On the wallet side, 23 of the 36 discoverable EVM-compatible wallets expose their provider interface into embedded iframes, covering 27.76 million users (78.95\% of the entire \NewEightyFive{} dataset). Any such wallet automatically hands its provider to whatever dApp the malicious site embeds, without user interaction or visibility.

Figure~\ref{fig:user-decision} shows that almost all realistic user workflows expose the user’s address to third-party trackers. If the user never disconnects from the dApp (or can't), and the wallet is unlocked, \textbf{all these 23 wallets leak the user's active address immediately}. When users do disconnect, in many cases, the problem persists. Among these 23 iframe-exposing wallets: 

\begin{itemize}
    \item \textbf{16 are revocation-unsafe}: even if the user wants to disconnect from the dApp side, the user is unable to. 
    \item \textbf{14 leaks the address even when locked}: 69\% of users are still under threat even when the wallet is locked.
\end{itemize}

In practice, for a very large portion of users, locking does not protect them, disconnecting does not protect them, and wallet revocation is simply not implemented. For millions of users, \textbf{wallet address leakage is just the default outcome.}

\subsection{\textbf{Discussion}}

Web-side linkability arises from inconsistent permission handling, iframe exposure, and widespread third-party scripts across wallets and dApps, enabling passive address recovery across sessions and sites even when users believe they have disconnected or locked their wallet. Taken together, the results show that these leaks are not isolated bugs but systemic interactions across wallets, dApps, ecosystem standards, and embedded scripts. Because the permission model is inconsistently enforced, and because iframe exposure bypasses traditional Web2 tracking defenses, \textbf{a determined tracker can link user activity across the web and even connect wallet addresses to a user’s real-world identity despite careful cookie or storage hygiene.}




\section{Summary and Mitigation}\label{mitigation}
Table~\ref{tab:threat-summary} summarizes the five privacy threats we identified in browser-extension wallets, their causes, and their ecosystem impact. We now outline several mitigations that can reduce these privacy risks.

\paragraph{\textbf{Network-Threat\#1}} Wallets should \textbf{avoid batching multiple user addresses into a single RPC request}, as this creates strong address co-occurrence linkability signals without providing functional benefits beyond developer convenience. Each RPC request should contain only one address to prevent direct linkage. Additionally, timing obfuscation (e.g., randomized delays or lightweight mixing) can be applied on the user side to reduce time-correlation linkability.

\paragraph{\textbf{Web-Threat\#1}} 
One possible way to prevent fingerprinting is to block EIP-6963 discovery events. However, wallet extensions may inject page scripts and dispatch these events before any user-side mechanism can intervene, making this defense unreliable. A more effective strategy would be for wallets to only enable announce messages when the wallet is unlocked. This allows dApps to detect wallets only when the user chooses to do so, while preventing unsolicited exposure of the user’s installed-wallet set.

\paragraph{\textbf{Web-Threats \#2--\#3}}
These threats originate from long-lived stale permissions stored by wallet extensions. We observed that this is not solely a wallet-side problem, but an ecosystem-wide issue, as most dApps also fail to revoke permissions properly.

We therefore propose three mitigations:

\textbf{(1) Ecosystem-wide revocation semantics.}
 Wallet address permission revocation should be defined and implemented by both wallets and dApps. 
 
\textbf{(2) Permission expiry}: Wallets should limit the lifetime of address permissions to avoid the existence of long-lived stale permissions.
Brave Wallet already adopts this approach. 

\textbf{(3) Restrict repeated account access}: Wallets should not allow dApps to repeatedly query address information without additional user interaction. Subsequent queries should require explicit consent from the user, preventing continuous passive access by the dApp.

However, these measures still do \emph{not} eliminate the tracking risk during the time window in which a permission remains valid, allowing a tracker embedded in the web page to still access it. 

To address it, we propose a newly designed \textbf{script-level access-controlled localStorage} for wallet permissions, similar to prior work on per-script-domain cookie isolation~\cite{cookieguardNikkhah}. Current localStorage is bound only to the page origin, meaning that third-party scripts embedded in the web page share the same page origin and can access the same data. Our design adds a check if the origin of the calling script matches the origin of the web page, thereby preventing embedded third-party scripts from accessing it. In this design, the wallet stores its per-origin permissions in script-level access-controlled localStorage rather than in extension-controlled storage, so that only first-party page scripts can access the associated addresses. An additional benefit is that permissions stored in localStorage are automatically cleared when users clear site data. Although browsers do not currently provide such script-level storage isolation natively, wallet extensions can approximate this design today by performing the same script-origin check based on the JavaScript Stack before returning addresses.

\paragraph{\textbf{Web-Threat \#4}}
This threat arises from wallet providers being injected into cross-origin contexts. The only reliable mitigation is to restrict provider exposure.

Wallets should therefore \textbf{inject providers only into top-level, same-origin contexts}. In addition, dApps should enforce standard framing protections (e.g., frame-ancestors or X-Frame-Options) to prevent malicious embedding.

Alternatively, wallets \textbf{may continue injecting providers into cross-origin contexts but enforce strict origin verification before serving requests}. Specifically, the wallet verifies that the origin of the calling script matches the top-level page origin before returning any wallet data. 


\section{Related Work}~\label{sec:relatedwork}

Prior research has extensively examined web-based device fingerprinting and browser-exposed identifiers~\cite{nick2013cookieless, vastel2018fp, pugliese2020long, gomez2018hiding}. Work such as Carnus~\cite{karami2020carnus} shows that browser extensions leak fingerprintable artifacts, while systems like CloakX~\cite{Erik2019CloakX} and Simulacrum~\cite{soroush2022Simulacrum} attempt to mask extension-induced signals. Agarwal et al.~\cite{Shubham2024peekingthroughthewindow} further show that extension code injection and event-driven behaviors are observable from the page. 
Our work focuses specifically on \textbf{Web3 wallet extensions} and highlights how emerging wallet standards introduce \textbf{new, Web3-native fingerprinting vectors}.

A separate line of work studies 
off-chain network-layer or on-chain metadata leakage. Users can be deanonymized via P2P traffic patterns~\cite{alex2015bitcoinp2p,biryukov2019deanonymization,heimbach2025deanonymizing}, RPC timing correlations~\cite{shan2025timetellall,Wang2024DeanonymizingEU}, centralized RPC infrastructure~\cite{kailun2023Badapples}, and on-chain graph analysis~\cite{androulaki2013evaluatingbitcoinprivacy,heidelberg2013bitcoin, victor2020ethereumaddress, sarah2016bitcoinpaymentswithnames,ting2018Etheruem,ferenc2021blocckhainwatching,KapposAnonymityInZcash}. These approaches focus on blockchain-level or network-level signals rather than the \textbf{wallet extension}, which is often the user’s first point of contact with Web3.

Within the browser-extension wallet ecosystem, prior work has mostly examined their security. Houy et al.~\cite{houy2023securityWallet} and recent SoKs~\cite{erinle2025sok} systematize wallet architectures and high-level attack surfaces. WalletRadar~\cite{Xia2024walletRadar} performs large-scale analysis to uncover code-level vulnerabilities but does not study privacy leakage through web- or network-side behaviors. Our work fills this gap.

Closest to our study are Torres et al.~\cite{torres2023walletprivacy} and Winter et al.~\cite{winter2023securityprivacydecentralizationweb3}. Torres et al.~\cite{torres2023walletprivacy} measure the endpoints contacted by wallets but treat all requests uniformly. In practice, many of these requests are functional RPC calls; we instead analyze the \textbf{privacy implications of multi-address exposure within these RPC flows}, which their endpoint-level analysis does not capture. Winter et al.~\cite{winter2023securityprivacydecentralizationweb3} show that DeFi frontends embed third-party trackers and frequently leak wallet addresses to analytics providers, enabling cross-site and Personally Identifiable Information (PII) linkage. Our results reveal a related but distinct problem: a tracker can learn a user’s address without any action from the dApp, because \textbf{wallets themselves passively disclose addresses}. This issue enables cross-site tracking, account clustering, and deanonymization
Prior work has mostly reported on dApp leaks or network metadata in isolation. Our results show that modern Web3 privacy risks emerge from the \textbf{interaction between wallets, dApps, third-party endpoints, and new wallet standards}, revealing an ecosystem-level attack surface not previously characterized.

\section{Conclusion}~\label{sec:Conclusion}

We identify five concrete privacy threats of browser-extension wallets and quantify their prevalence using measurement frameworks that we developed to capture both network-side and web-side traffic. Our findings show that a large majority of active browser-extension wallet users are affected by these threats. 

In particular, a central cause of these threats is the mismatch between wallet capabilities and their privacy guarantees. Wallets allow users to manage multiple addresses within a single interface. Yet, on the network side, wallets often batch or structure network requests in ways that reveal relationships between a user's addresses. On the web side, wallet discovery APIs, the lack of ecosystem-wide revocation mechanisms, and cross-origin provider injection expose information about users and their addresses in ways that users may not expect.

As a result, these behaviors enable external adversaries to infer multi-address ownership, fingerprint Web3 wallet users, enable persistent user tracking across sessions and sites, and ultimately link users’ web activity to their on-chain wealth. 

We discuss several mitigations for these threats. Addressing them will require more privacy-preserving handling of wallet-related network requests, stronger privacy considerations in ecosystem standards for wallet discovery and permission revocation, and stricter controls over cross-origin provider exposure to place user privacy at the core of Web3 wallet design.



\section{Open Source and Demo}\label{sec:EthicalPrinciples}

Our frameworks, datasets, and analysis code are released as open source\footnote{
Artifact repository: \url{https://github.com/podiumdesu/wallet-privacy-threats}. }, and we provide a public demo that allows anyone to easily test the web-side threats of wallet extensions.\footnote{Web exposure demo: \url{https://wallet-privacy.distriled.dnetcloud.cs.kuleuven.be/}.
}

\section{Responsible Disclosure}\label{sec:responsible}

We focused our disclosure efforts on web-side threat \#4, which does not stem from ecosystem standards and can be addressed directly by wallet developers. The other threats primarily arise from ecosystem-level design decisions or wallet implementation choices.

In February 2026, we retested the affected wallets using the latest versions available on the Chrome Web Store with our web exposure demo\footnotemark[5]. Out of the 23 wallets originally affected by threat \#4, two wallets (Coinbase Wallet v3.120.0 and Coin98 v10.4.1) were no longer vulnerable in their latest versions. We therefore contacted the vendors of the remaining 21 wallets. 

Within a one-month notification window prior to the camera-ready submission, eight wallets responded to our reports through their bug bounty programs, including MetaMask, OKX, Trust, Rabby, Backpack, Bybit, Zerion, and Core. Most vendors simply classified the issue as informational or out of scope for their bug bounty programs. 

OKX acknowledged the technical correctness of our findings and the associated privacy implications, but classified the issue as informational because it lacks demonstrable functional or financial harm beyond information disclosure.

MetaMask confirmed that cross-origin provider exposure is a known risk internally and was one of the main motivations behind their development of an alternative wallet API that does not rely on provider injection. They further stated that they currently have no immediate plans to stop injecting the provider, as doing so would create significant breaking changes for the dApp ecosystem, although future changes may be considered.

\begin{acks}
The authors used generative AI–based tools (ChatGPT) to revise the text, improve clarity and precision, correct grammatical errors, and smooth out awkward phrasing. ChatGPT was also used to generate parts of the analysis code, which were carefully reviewed and validated by the authors.

We thank the anonymous reviewers of PoPETs for their constructive and insightful feedback, which helped improve the clarity and quality of this paper. This research was partially supported by the Research Fund KU Leuven and the Cybersecurity Research Program Flanders.
\end{acks}
\bibliographystyle{ACM-Reference-Format}
\bibliography{references}

\newpage
\appendix

\section{Analytics sites list}\label{appendix:analytics-list}

This appendix provides the full list of 21 third-party analytics and tracking domains used in our classification, grouped by provider category.
\\
\begin{center}
\small
\begin{tabular}{ll}
\toprule
\textbf{Category} & \textbf{Domain} \\
\midrule
Google Analytics / Ads & google-analytics.com \\
 & googletagmanager.com \\
 & analytics.google.com \\
 & g.doubleclick.net \\
 & stats.g.doubleclick.net \\
 & doubleclick.net \\
 & googletagservices.com \\
\midrule
Product Analytics & segment.io \\
 & amplitude.com \\
 & mixpanel.com \\
\midrule
Error / Telemetry & sentry.io \\
 & bugsnag.com \\
 & newrelic.com \\
 & datadoghq.com \\
\midrule
User Engagement & intercom.io \\
 & intercomcdn.com \\
 & hotjar.com \\
 & fullstory.com \\
\midrule
A/B Testing & optimizely.com \\
\midrule
Other Trackers & clarity.ms \\
 & facebook.com \\
\bottomrule
\end{tabular}
\end{center}

\section{Observed Node Providers and RPC endpoints}
\label{appendix:rpc-providers}

Wallets query external node providers and RPC endpoints to retrieve blockchain state (e.g., balances, nonces, and transaction metadata). These services include commercial node providers, chain-operated infrastructure, and blockchain data APIs. Examples observed in our measurements include.
\\
\begin{center}
\small
\begin{tabular}{ll}
\toprule
\textbf{Category} & \textbf{Domain} \\
\midrule
Commercial Node Providers & infura.io \\
 & ankr.com \\
 & drpc.org \\
 & publicnode.com \\
 & 1rpc.io \\
\midrule
Chain Infrastructure & arbitrum.io \\
 & avax.network \\
 & binance.org \\
\midrule
Blockchain Data APIs & etherscan.io \\
 & aptoslabs.com \\
\bottomrule
\end{tabular}
\end{center}

These domains are representative examples observed in our measurements and do not constitute an exhaustive list of infrastructure services contacted by wallets.\footnote{A larger community-maintained list of RPC node providers is available at \url{https://github.com/arddluma/awesome-list-rpc-nodes-providers}.}

\end{document}